# Ultrahigh-temperature ferromagnetism in ultrathin insulating films with ripple-infinite-layer structure

Yazhuo Yi[1#], Haoliang Huang[2,3#], Ruiwen Shao[4#], Yukuai Liu[5#], Guangzheng Chen[1], Jiahui Ou[1], Xi Zhang[6], Ze Hua[3], Lang Chen[2], Chi Wah Leung[7], Xie-Rong Zeng[1], Feng Rao[1], Nan Liu[8], Heng Wang[2], Liang Si[8,9*], Hongyu An[10*], Zhuoyu Chen[2,3*], Chuanwei Huang[1*]

1. Shenzhen Key Laboratory of Special Functional Materials, Guangdong Provincial Key Laboratory of New Energy Materials Service Safety, College of Materials Science and Engineering, Shenzhen University, Shenzhen 518060, China

2. Department of Physics, Southern University of Science and Technology, Shenzhen 518055, China

3. Quantum Science Center of Guangdong-Hong Kong-Macao Greater Bay Area, Shenzhen 518045, China

4. Beijing Advanced Innovation Center for Intelligent Robots and Systems, School of Medical Technology, Beijing Institute of Technology, Beijing 100081, P. R. China.

5. College of Electronic Information and Mechatronic Engineering, Zhaoqing University, Zhaoqing Road, Duanzhou District, Zhaoqing, 526061, China

6. Electron Microscopy Centre of Shenzhen University, Shenzhen University, Shenzhen, 518060, China

7. Department of Applied Physics, The Hong Kong Polytechnic University, Hung Hom, Hong Kong, China

8. School of Physics, Northwest University, Xi'an 710127, China

9. Institute of Solid State Physics, TU Wien, 1040 Vienna, Austria

10. College of New Materials and New Energies, Shenzhen Technology University, Shenzhen 518118, China

________________________________________________________________

#These authors contributed equally.

*E-mails: siliang@nwu.edu.cn, anhongyu@sztu.edu.cn, chenzhuoyu@sustech.edu.cn, cwhuang@szu.edu.cn

**Abstract**

Ferromagnetism and electrical insulation are often at odds, signifying an inherent trade-off. The simultaneous optimization of both in one material, essential for advancing spintronics and topological electronics, necessitates the individual manipulation over various degrees of freedom of strongly correlated electrons. Here, by selective control of the spin exchange and Coulomb interactions, we report the achievement of $SrFeO_2$ thin films with resistivity above $10^6$ Ω·cm and strong magnetization with Curie temperature extrapolated to be ~ 1200 K. Robust ferromagnetism is obtained down to ~1.0 nm thickness on substrate and ~ 2.0 nm for freestanding films. Featuring an out-of-plane oriented ripple-infinite-layer structure, this ferromagnetic insulating phase is obtained through extensive reduction of as-grown brownmillerite $SrFeO_{2.5}$ films at high compressive strains. Pronounced spin Hall magnetoresistance signals up to ~ 2.6‰ is further demonstrated with a Pt Hall-bar device. Our findings promise emerging spintronic and topological electronic functionalities harnessing spin dynamics with minimized power dissipations.

Despite the pivotal need in spintronic and topological electronic technologies (*1-3*), the quest for materials that exhibit both ferromagnetism and electrical insulation presents a formidable challenge(*4-6*). Ferromagnetism is usually based upon the presence of partially filled *d*- or *f*-orbitals, where electrons with unpaired electrons can engage in exchange interactions. In contrast, insulation in band theory requires fully occupied electron orbitals, thereby precluding the free movement that facilitates magnetism. This intricate interplay of electronic structure and charge carrier dynamics reflects the fundamental trade-off that defines the pursuit of ferromagnetic insulators (FMIs)(*7*).

To conquer the ferromagnetism-insulation trade-off, innovative material design strategies that leverage strong electron correlations and multiple electron orbitals are essential. However, the search for such materials is complicated by indirect interactions like super-exchange and the Ruderman–Kittel–Kasuya–Yosida (RKKY) coupling(*8-10*). Therefore, most ferromagnetic insulators identified to date exhibit Curie temperatures ($T_C$) significantly below room temperature(*4, 5, 11-17*), limiting their viability for practical device applications. While ferrimagnetism with counterbalancing of opposing spins may offer higher $T_C$ surpassing room temperature(*18-20*), the simultaneous achievement of elevated ferromagnetic $T_C$ and adequately high resistivity in a single material remains to be a task to the field.

Practically functional FMIs operating above room temperature represent the cornerstones for future spintronic and topological electronic innovations(*21, 22*). For instance, FMI is indispensable for introducing robust magnetic order into topological systems without disrupting their quantum protected states, including the dissipationless spin-polarized edge states in quantum anomalous Hall effect (QAHE) systems (*3, 23, 24*). Ultrathin FMIs also serve as critical components in magnetic tunnel junctions (MTJ) for non-volatile memory applications (*25*), and in spin valves to enhance spin-filtering capabilities and reduce spin damping, for advanced spintronic operations (*26*). To realize device integrations, the compatibility of FMIs with a variety of substrates (*27, 28*), especially silicon (*29, 30*), is important. However, the majority of FMIs are typically found in the form of bulk crystals or as epitaxial thin films on rigid substrates. In addition, for advancing spintronic applications, the performance of spin transport in FMI-integrated devices is a primary concern, particularly for spin transport at the interface between the FMI and heavy metals (*31, 32*).

In this study, we present an exotic FMI phase of $SrFeO_2$, characterized by the infinite stacking of rippling 4-coordinate $FeO_2$ layers. Markedly different from typical infinite-layer (IL) $SrFeO_2$ with $FeO_2$ planes parallel to substrate surface (Fig. 1A, left) (*33-36*), this ripple-infinite-layer (RIL) $SrFeO_2$ is featured by out-of-plane oriented rippling $FeO_2$ planes (Fig. 1A, right). Spin-polarized density-functional theory plus $U$ (DFT+$U$) calculations underscore the dichotomy in energetic preferences between these two phases, demonstrating inclinations for antiferromagnetism (AFM) in the IL phase and ferromagnetism (FM) in the RIL phase (more details are shown in fig. S1-2 and table S1).

The occurrence of RIL phase is highly sensitive to the misfit strain $\varepsilon$ imparting from the underlying substrate (Fig. 1B and fig. S3 to S5). The RIL phase is obtained through an extensive $CaH_2$ topotactic reduction process exceeding 120 hours of brownmillerite (BM) $SrFeO_{2.5}$ films grown on $YAlO_3$ substrates that imposes substantial compressive strains over 6%. Intriguingly, two mixed-phase regimes (i.e., BM & IL and IL & RIL, shadowed regions in fig. S5E-G) exist during the reduction. Noted that no RIL phase can be formed in overly thick films (i.e. $\gtrsim$ 50 nm) even after prolonged reduction (> 220 hours, fig. S6). In cases with lower strain, only the IL phase is observed after reduction, consistent with previously reported results (*34, 37*). The structural distinctions between the IL and RIL phases are clearly resolved in X-ray diffractions (XRD) (Fig. 1C). X-ray absorption spectra (XAS) show clear differences among Fe L-edges of the BM, IL and RIL phases, exhibiting a valence transition from $Fe^{3+}$ to $Fe^{2+}$ (Fig. 1D), accompanied by an enhancement of Fe-O hybridization (shadowed region in Fig.1E). Importantly, the XAS results do not show substantial differences of electronic structure between IL and RIL phases.

Scanning transmission electron microscopy (STEM) were conducted to differentiate the microcrystalline structures of the $SrFeO_{3-\delta}$ (SFO, $0.5 \leq \delta \leq 1.0$) with different BM, IL, and RIL phases. The lattice constants are clearly distinguished using high-angle annular dark-field (HAADF) STEM (fig.S7, B-G), aligning well with the XRD results. The annular bright-field (ABF)-STEM images (fig.S7, H-J) reveal distinct coordinate frameworks. An integrated differential phase contrast (iDPC) method was employed to confirm the microstructures of the RIL phase across a large field of view (Fig. 1F) (*38*). The vertically oriented $FeO_2$ layers exhibit an apparent rippling structure (Fig. 1G). Interestingly, the Fe-O bonds parallel to the substrate interface alternate between straight (Fig. 1H-I) and buckling geometries (Fig. 1J-K), revealing a highly complex interplay of structural features. The synergistic analysis of STEM and XAS data reveals that the distinct magnetic properties of the IL and RIL phases are predominantly attributed to the structural deformation within the $FeO_2$ layers, rather than alterations in the valence state or oxygen hybridization. Specifically, the observed deformations of the Fe-O bond angle may reduce the Fe-O-Fe super-exchange interaction and simultaneously enhance the Fe-Fe direct exchange interaction, favoring ferromagnetic state. Importantly, despite rippling deformations, the 4-coordinate intralayer geometry is retained as in the IL structure.

The iDPC images also reveal a significant displacement of $Sr^{2+}$ ions (~1.40 Å as shown in fig. S8) between alternating Sr-O layers along the [110] direction. This concurrently leads to a substantial distortion of the $FeO_4$ coordinate between the two displaced Sr-O layers, as quantitatively depicted in Fig. 1G. Additionally, energy dispersive spectroscopy (EDS) analysis (fig. S9) reveals sharp chemical interfaces and uniform elemental distribution profiles (such as Sr, Fe, and O) for the reduced RIL film, effectively ruling out elemental decomposition of $SrFeO_2$ films during the reduction treatment. Furthermore, time-of-flight secondary-ion mass spectrometry (TOF-SIMS) measurements show no evident signs of hydrogen intercalation (fig. S10), thus excluding unexpected effects from hydrogen insertions, unlike structures reported in the nickelates (*39, 40*).

The RIL films demonstrate a desirable magnetization of approximately 180 emu/cc at low temperatures (Fig. 2A), surpassing most FMIs reported to date (*4, 6, 11, 13, 41-43*). Remarkably, the RIL phase maintains a robust ferromagnetism at temperatures approaching 900 K, the point at which the RIL phase begins to be partially re-oxidized (fig. S12). While the structural phase transition temperature is primarily determined by the thermodynamic stability, the Curie temperature $T_C$ is defined to correlate with the strength of ferromagnetic interactions. As an indicator of the robustness of ferromagnetism, we extrapolate the magnetization-temperature curve to yield an estimated $T_C$ of around 1200 K. This ultrahigh $T_C$ is reconfirmed in the mixed IL & RIL phase of the reduced SFO/LAO (or SFO/LASO) system (fig. S11). Figure 2B illustrates the temperature dependent hysteresis loops of the 20-nm thick RIL film, exhibiting a restorable hysteresis loop up to 750 K (fig. S12C). At different reduction stages (BM-IL-RIL), magnetic performances are systematically characterized (fig. S13 and S14). In RIL films, hysteresis loops are detectable as thin as 1 nm even at 400 K (Fig. 2C). Ultrathin FMI films ($\lesssim$ 2 nm) with $T_C$ above room temperature, as an indispensable materials foundation for the spintronic devices, are rarely found previously (*25*).

The RIL films can be further lifted off from substrates by etching away a 2-nm water-soluble sacrificial buffer layer $Sr_3Al_2O_6$ (SAO) grown prior (*27*). To shield the RIL films from oxygen absorption or extraction, the pristine BM $SrFeO_{2.5}$ is encapsulated by 2-nm oxygen-deficient LaAlO$_{3-\delta}$ layers during film growth. The RIL-SFO films can be readily transferred from the rigid perovskite substrate YAlO$_3$ to various desired substrates, such as PET, silicon, or glass, while preserving its crystalline structure (Fig. 2E), smooth surface morphology (fig. S15), and magnetic properties (Fig. 2F and fig. S16). Notably, in contrast to other strain-induced properties, the characteristic XRD peaks of the transferred RIL (~ 46° in Fig. 2E) is well-maintained even after the release and transfer process. Furthermore, the transferred RIL-SFO films, detached from the YAO substrate, retains robust ferromagnetic properties (with minor weakening) with a thickness down to 2 nm at both 10 K and 400 K (Fig. 2F). These findings indicate the superior robustness of the ferromagnetism of RIL-SFO, facilitating the integration into flexible or silicon-based spintronic device platforms.

Given the intrinsic trade-off that exists between ferromagnetism and electrical insulation, it is typically challenging to achieve both a high $T_C$ and high resistivity within a single-phase material. As shown in Fig. 3A, optical absorption measurement reveals that the RIL phase has large band gap of around 3.27 eV, similar to the measured band gap of the IL phase (3.45 eV) and the mixed IL & RIL phase (3.29 eV). Despite marked change in magnetic properties, the RIL phase retains the electrical insulation typical of the IL. Fig. 3B compiles data on representative ferromagnetic and ferrimagnetic insulators, broadly demonstrating an inverse relationship between $T_C$ and resistivity. Notably, the RIL SrFeO$_2$ films, while displaying remarkably strong ferromagnetism, also possess exceptional insulating characteristics. Its resistivity has gone beyond the maximum limit of our measurement setup (see Methods), such that we conclude that the resistivity is greater than $10^6$ Ω·cm, in sharp contrast to the resistivity measured in weakly constrained SrFeO$_2$ (fig. S17). The simultaneous achievement of high $T_C$ of around 1200 K and a high resistivity over $10^6$ Ω·cm transcends beyond the fundamental ferromagnetism-electrical insulation trade-off, highlighting a significant advancement in the search for FMI.

We further demonstrate spin transport properties in heavy metal/FMI Hall bar devices (Fig. 4). The devices are fabricated by patterning platinum Hall bars directly on top of the RIL (and BM for reference) thin films (inset in Fig. 4C). We quantified the spin Hall magnetoresistance (SMR) effect with rotating magnetic field as a function of film thickness (Fig. 4A). At 300 K, the Pt/RIL device exhibits significantly improved spin transport performance over the Pt/BM device. The highest magnetoresistance (MR) ratio ($\Delta R_{xx}/R_{xx,0}$), recorded at 300 K for a 10-nm-thick RIL film (Fig. 4B), is around 2.6‰. Among a range of heavy metal/ferromagnet and heavy metal/ferrimagnet devices, the Pt/RIL devices exhibit a remarkable enhancement of over 200% in performance compared to previous state-of-the-arts (Fig. 4C) (*31, 32, 44-51*). These results indicate that the RIL film, with its ultralow electrical conductivity, elevated $T_C$, and significant SMR, is an exemplary ferromagnetic insulating candidate for spintronic applications. Such devices, promising ultralow energy dissipation and high-speed operation, are capable of efficiently generating and conveying pure spin currents across a broad temperature spectrum.

In conclusion, this study uncovers the ripple-infinite-layer $SrFeO_2$ ferromagnetic insulating films, with >1000 K Curie temperature, >3 eV insulating bandgap, >$10^6$ Ω·cm resistivity, >2.5‰ spin Hall magnetoresistance, and the flexibility to be transferred to different substrates. The attainment of FMI, overcoming the fundamental ferromagnetism-electrical insulation trade-off by rippling the $FeO_2$ layers, represents a selective control of spin and charge degrees of freedom in strongly correlated electronic materials. This discovery not only facilitates spintronic and topological electronic applications, but also opens new avenues for designing novel extreme properties in correlated electronic materials.

## References

1. S. Manipatruni *et al.*, Scalable energy-efficient magnetoelectric spin-orbit logic. *Nature* **565**, 35-42 (2018).

2. H. Yang *et al.*, Two-dimensional materials prospects for non-volatile spintronic memories. *Nature* **606**, 663-673 (2022).

3. K. He, Y. Wang, Q.-K. Xue, Topological Materials: Quantum Anomalous Hall System. *Annual Review of Condensed Matter Physics* **9**, 329-344 (2018).

4. D. Meng *et al.*, Strain-induced high-temperature perovskite ferromagnetic insulator. *Proceedings of the National Academy of Sciences* **12**, 2873 (2018).

5. A. Schmehl *et al.*, Epitaxial integration of the highly spin-polarized ferromagnetic semiconductor EuO with silicon and GaN. *Nat. Mater.* **6**, 882-887 (2007).

6. Q. Zhang *et al.*, Near-room temperature ferromagnetic insulating state in highly distorted $LaCoO_{2.5}$ with $CoO_5$ square pyramids. *Nat Commun* **12**, 1853 (2021).

7. Y. Matsumoto *et al.*, Room-Temperature Ferromagnetism in Transparent Transition Metal-Doped Titanium Dioxide. *Science* **291**, 854-856 (2001).

8. M. A. Ruderman, C. Kittel, Indirect Exchange Coupling of Nuclear Magnetic Moments by Conduction Electrons. *Phys. Rev.* **96**, 99-102 (1954).

9. K. Yosida, Magnetic Properties of Cu-Mn Alloys. *Phys. Rev.* **106**, 893-898 (1957).

10. T. Kasuya, A Theory of Metallic Ferro- and Antiferromagnetism on Zener's Model. *Prog. Theor. Phys.* **16**, 45-57 (1956).

11. H. Boschker *et al.*, High-Temperature Magnetic Insulating Phase in Ultrathin LSMO Films. *Phys. Rev. Lett.* **109**, 157207 (2012).

12. A. Sadoc *et al.*, Large Increase of the Curie Temperature by Orbital Ordering Control. *Phys. Rev. Lett.* **104**, 046804 (2010).

13. M. Gajek *et al.*, Spin filtering through ferromagneticBiMnO3tunnel barriers. *Phys. Rev. B* **72**, 020406 (2005).

14. S. Chakraverty *et al.*, $BaFeO_3$ cubic single crystalline thin film: A ferromagnetic insulator. *Appl. Phys. Lett.* **103**, 142416 (2013).

15. J. H. Lee *et al.*, A strong ferroelectric ferromagnet created by means of spin-

lattice coupling. *Nature* **466**, 954-958 (2010).

16. J. S. Moodera, X. Hao, G. A. Gibson, R. Meservey, Electron-Spin Polarization in Tunnel Junctions in Zero Applied Field with Ferromagnetic EuS Barriers. *Phys. Rev. Lett.* **61**, 637-640 (1988).

17. L. F. Kourkoutis, J. H. Song, H. Y. Hwang, D. A. Muller, Microscopic origins for stabilizing room-temperature ferromagnetism in ultrathin manganite layers. *Proc. Natl. Acad. Sci. U. S. A.* **107**, 11682-11685 (2010).

18. X. Chen *et al.*, Magnetotransport Anomaly in Room-Temperature Ferrimagnetic $NiCo_2O_4$ Thin Films. *Adv. Mater.* **31**, 1805260 (2019).

19. U. Lüders *et al.*, $NiFe_2O_4$: A Versatile Spinel Material Brings New Opportunities for Spintronics. *Adv. Mater.* **18**, 1733-1736 (2006).

20. A. Mitra *et al.*, Interfacial Origin of the Magnetisation Suppression of Thin Film Yttrium Iron Garnet. *Scientific Reports* **7**, 11774 (2017).

21. S. Salahuddin, K. Ni, S. Datta, The era of hyper-scaling in electronics. *Nature Electronics* **1**, 442-450 (2018).

22. J. Puebla, J. Kim, K. Kondou, Y. Otani, Spintronic devices for energy-efficient data storage and energy harvesting. *Communications Materials* **1**, 24 (2020).

23. C.-Z. Chang, C.-X. Liu, A. H. MacDonald, Colloquium: Quantum anomalous Hall effect. *Reviews of Modern Physics* **95**, 0110002 (2024).

24. C. Tang *et al.*, Above 400-K robust perpendicular ferromagnetic phase in a topological insulator. *Science Advances* **3**, e1700307 (2017).

25. M. Gajek *et al.*, Tunnel junctions with multiferroic barriers. *Nat. Mater.* **6**, 296-302 (2007).

26. W. Han, S. Maekawa, X.-C. Xie, Spin current as a probe of quantum materials. *Nat. Mater.* **19**, 139-152 (2019).

27. D. Lu *et al.*, Synthesis of freestanding single-crystal perovskite films and heterostructures by etching of sacrificial water-soluble layers. *Nat. Mater.* **15**, 1255-1260 (2016).

28. J. Zhang *et al.*, Super-tetragonal $Sr_4Al_2O_7$ as a sacrificial layer for high-

integrity freestanding oxide membranes. *Science* **383**, 388-394 (2024).

29. L. Han *et al.*, High-density switchable skyrmion-like polar nanodomains integrated on silicon. *Nature* **603**, 63-67 (2022).

30. S. S. Cheema *et al.*, Enhanced ferroelectricity in ultrathin films grown directly on silicon. *Nature* **580**, 478-482 (2020).

31. C. O. Avci *et al.*, Current-induced switching in a magnetic insulator. *Nat. Mater.* **16**, 309-314 (2017).

32. B. F. Miao, S. Y. Huang, D. Qu, C. L. Chien, Physical Origins of the New Magnetoresistance in Pt/YIG. *Phys. Rev. Lett.* **112**, 236601 (2014).

33. C. Tassel, H. Kageyama, Square planar coordinate iron oxides. *Chem. Soc. Rev.* **41**, 2025-2035 (2012).

34. Y. Tsujimoto *et al.*, Infinite-layer iron oxide with a square-planar coordination. *Nature* **450**, 1062-1065 (2007).

35. D. Li *et al.*, Superconductivity in an infinite-layer nickelate. *Nature* **572**, 624-627 (2019).

36. W. J. Kim *et al.*, Geometric frustration of Jahn-Teller order in the infinite-layer lattice. *Nature* **615**, 237-243 (2023).

37. S. Inoue *et al.*, Anisotropic oxygen diffusion at low temperature in perovskite-structure iron oxides. *Nat Chem* **2**, 213-217 (2010).

38. P. Nukala *et al.*, Reversible oxygen migration and phase transitions in hafnia-based ferroelectric devices. *Science* **372**, 630-635 (2021).

39. L. Si *et al.*, Topotactic Hydrogen in Nickelate Superconductors and Akin Infinite-Layer Oxides $ABO_2$. *Phys. Rev. Lett.* **124**, 166402 (2020).

40. X. Ding *et al.*, Critical role of hydrogen for superconductivity in nickelates. *Nature* **615**, 50-55 (2023).

41. Y. Wang *et al.*, Robust Ferromagnetism in Highly Strained $SrCoO_3$ Thin Films. *Physical Review X* **10**, 021030 (2020).

42. W. Li *et al.*, Interface Engineered Room-Temperature Ferromagnetic Insulating State in Ultrathin Manganite Films. *Advanced Science* **7**, 1901606

(2019).

43. Q. I. Yang *et al.*, Pulsed laser deposition of high-quality thin films of the insulating ferromagnet EuS. *Appl. Phys. Lett.* **104**, 082402 (2014).

44. S. Geprägs *et al.*, Static magnetic proximity effects and spin Hall magnetoresistance in Pt/$Y_3Fe_5O_{12}$ and inverted $Y_3Fe_5O_{12}$/Pt bilayers. *Phys. Rev. B* **102**, 214438 (2020).

45. A. Quindeau *et al.*, $Tm_3Fe_5O_{12}$/Pt Heterostructures with Perpendicular Magnetic Anisotropy for Spintronic Applications. *Advanced Electronic Materials* **3**, 1600376 (2016).

46. J. Q. Guo *et al.*, Spin Hall magnetoresistance of $CoFe_2O_4$/Pt heterostructures with interface non-collinear magnetic configurations. *Appl. Phys. Lett.* **121**, 142403 (2022).

47. K. Ganzhorn *et al.*, Spin Hall magnetoresistance in a canted ferrimagnet. *Phys. Rev. B* **94**, 094401 (2016).

48. Z. Ding *et al.*, Spin Hall magnetoresistance inPt/$Fe_3O_4$ thin films at room temperature. *Phys. Rev. B* **90**, 134424 (2014).

49. S. Crossley *et al.*, Ferromagnetic resonance of perpendicularly magnetized $Tm_3Fe_5O_{12}$/Pt heterostructures. *Appl. Phys. Lett.* **115**, 172402 (2019).

50. M. Althammer *et al.*, Role of interface quality for the spin Hall magnetoresistance in nickel ferrite thin films with bulk-like magnetic properties. *Appl. Phys. Lett.* **115**, 092403 (2019).

51. B.-W. Dong *et al.*, Spin Hall magnetoresistance in the non-collinear ferrimagnet GdIG close to the compensation temperature. *J. Phys.: Condens. Matter* **30**, 035802 (2018).

52. X. Guan, G. Zhou, W. Xue, Z. Quan, X. Xu, The investigation of giant magnetic moment in ultrathin $Fe_3O_4$ films. *APL Mater.* **4**, 036104 (2016).

53. M. Althammer *et al.*, Quantitative study of the spin Hall magnetoresistance in ferromagnetic insulator/normal metal hybrids. *Phys. Rev. B* **87**, 224401 (2013).

**Acknowledgements:**

R. S. acknowledges the Analysis and Testing Center at the Beijing Institute of Technology. X. Z. wish to acknowledge the assistance on FIB received from the Electron Microscope Center of the Shenzhen University. The authors thank the MCD-A beamline of the National Synchrotron Radiation Laboratory (NSRL) and the BL08U1A beamline of the Shanghai Synchrotron Radiation Facility (SSRF) for providing the XAS beam times. L.S. acknowledges support from the Vienna Scientific Cluster for the first-principles calculations. **Funding**: This work was financial supported by the Shenzhen Science and Technology Program (No.20220809153419002), the National Key R&D Program of China (Nos. 2022YFA1402903, 2022YFA1403100 and 2022YFA1403101), the National Natural Science Foundation of China (Nos. 12274025, 92265112, 12374455, 12274025, and 52032006), the Guangdong Provincial Quantum Science Strategic Initiative (Nos. GDZX2401004, GDZX2201001), the Shenzhen University 2035 Program for Excellent Research (No.00000203), the Guangdong Provincial Key Laboratory Program (No. 2021B1212040001) and Research Grants Council HKSAR (15302222).

**Author Contributions:**

C.H. initiated the project and orchestrated the experimental design. Y.Y., G.C., and J.O. were responsible for the fabrication of thin films, reduction treatment, and subsequent film release. Y.Y., Y.L., H.W., and H.H. conducted the X-ray diffraction, magnetic, and electrical measurements. N.L. and L.S. carried out the theoretical analysis and calculations. X.Z. and Z.H. performed the focused ion beam irradiation to create the TEM samples. Z.H. and R.S. executed the scanning transmission electron microscopy measurements. H.H. conducted the soft X-ray absorption measurements. H.A. carried out the spin Hall effect measurements. Z.C. and C.H. analyzed and interpreted the data. H.H., Y.L., F.R., L.C., C.L., X.Z., Z.C., and C.H. engaged in discussions about the results and wrote the manuscript. All authors provided comments on the manuscript.

**Competing interests:**

Authors declare that they have no competing interests

**Data and materials availability:**

All data are available in the main text or the supplementary materials.

**Supplementary Materials**

Materials and Methods

Figs. S1 to S17

Tables S1

References (*53–68*)

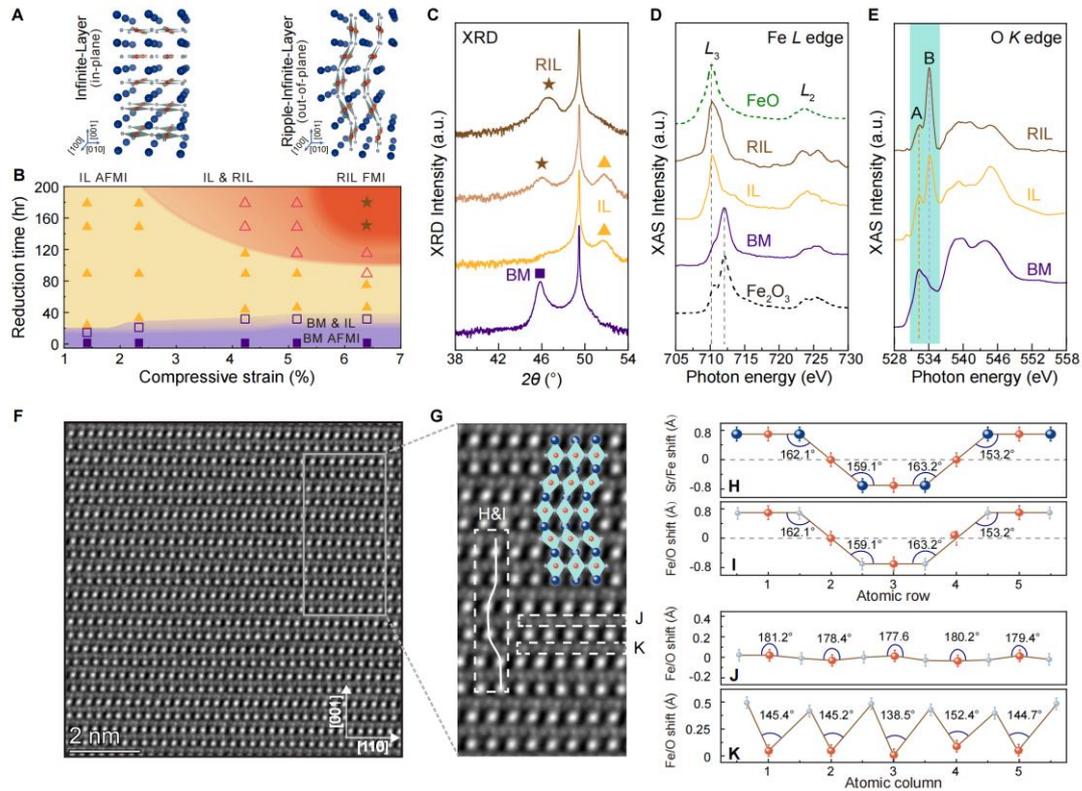

**Fig. 1.** Structural transformations of SrFeO$_{3-\delta}$ thin films. (**A**) Schematic crystal structures of IL-SFO with in-plane oriented Fe-O coordinates (left) and RIL-SFO with out-of-plane oriented Fe-O coordinates (right). (**B**) Synergistic effects of misfit strain and reduction reaction on the phase diagram of the SrFeO$_{3-\delta}$ film based on XRD results, consisting of pristine BM-SFO, IL-SFO and RIL-SFO. Hollow symbols present mixed BM&IL or IL&RIL phase. (**C**) X-ray diffraction measurements of the SrFeO$_{3-\delta}$/YAO heterostructure system, showing distinct crystalline structures of BM (square), to IL (triangle), then to mixed IL & RIL, and finally to RIL (asterisk) phase. (**D-E**) Corresponding X-ray absorption spectra of Fe L-edges and O K-edges for BM-SFO, IL-SFO and RIL-SFO, together with reference spectra from FeO and Fe$_2$O$_3$. The shifts in the peak positions of the Fe L-edge between BM-SFO and the other two reduced phases (i.e., IL-SFO and RIL-SFO) suggest a decrease in valence states from Fe$^{3+}$ to Fe$^{2+}$. The O K-edge results further show distinct differences (Peaks A and B in shadowed region) between BM-SFO and the other two reduced phases, indicating significant changes in Fe-O hybridization during the topotactic transformations. (**F-G**) Integrated differential phase contrast technique for imaging microstructure of the RIL-

SFO phase along the [110] direction. (**G**) Zoom-in image depicting distorted $FeO_4$ between two alternating SrO-SrO layers. (**H-I**) Periodic shifts of Sr/Fe ions, and corresponding angle changes of Sr-Fe-Sr (**H**) or O-Fe-O angle (**I**), indicated by the vertical curve in (**G**). (**J-K**) Relative shifts of Fe/O and distorted angles of O-Fe-O for two alternating FeO-FeO layers, showcasing distinct ion displacements and coordination distortions of the RIL-SFO phase along horizonal direction.

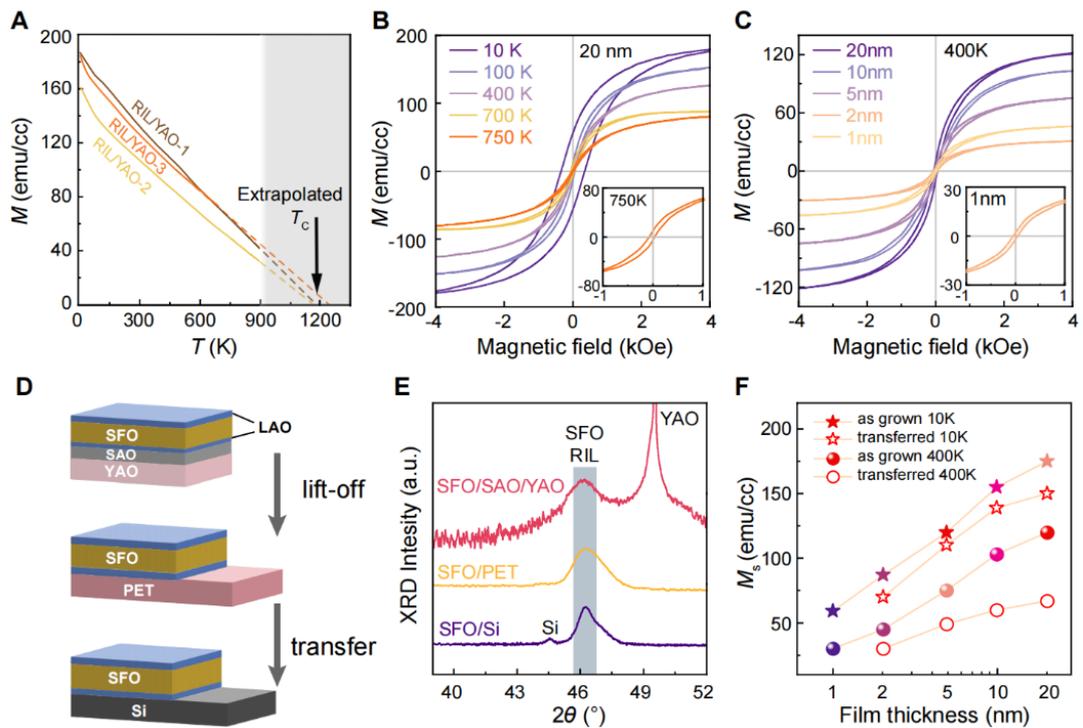

**Fig. 2.** Magnetic properties of the RIL-SFO film. (**A-B**) In-plane *M-T* curve and temperature-dependent *M-H* of a 20-nm-thick RIL-SFO film. H (~ 6000 Oe) was applied to the [100] direction. The $T_C$ value of RIL-SFO, estimated from the extrapolation of the $M-T$ curve, is up to 1200 K. (**C**) Thickness-dependent magnetic property of the RIL-SFO deposited on YAO. Inset shows a desirable ferromagnetic property of an ultrathin (~ 1.0 nm) RIL-SFO film. (**D**) Schematic representation of the release and transfer processes of the RIL-SFO film from the rigid YAO substrate to flexible PET, and then to the Si substrate. (**E**) Corresponding X-ray diffraction patterns of the constrained, released and transferred RIL-SFO films. (**F**) In-Plane $M_s$

of the constrained (solid points) and transferred (hollow points) RIL-SFO Films as a function of film thickness.

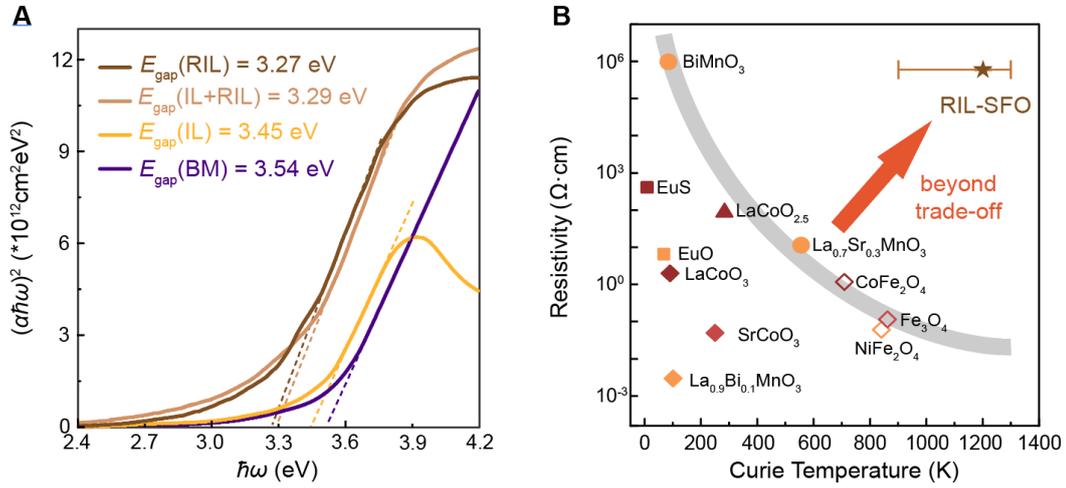

**Fig. 3.** (**A**) Optical absorption spectra for various phases of SrFeO$_{3-\delta}$ films. Through fitting of the optical absorption spectra, we estimated the bandgap of BM-SFO, IL-SFO and RIL-SFO, respectively. (**B**) The resistivity measured at 300 K and $T_c$ of the RIL-SFO film compared to the best-performing ferromagnetic (or ferrimagnetic, hollow symbols) insulators (*4-6, 11, 13, 18, 41-43, 52*).

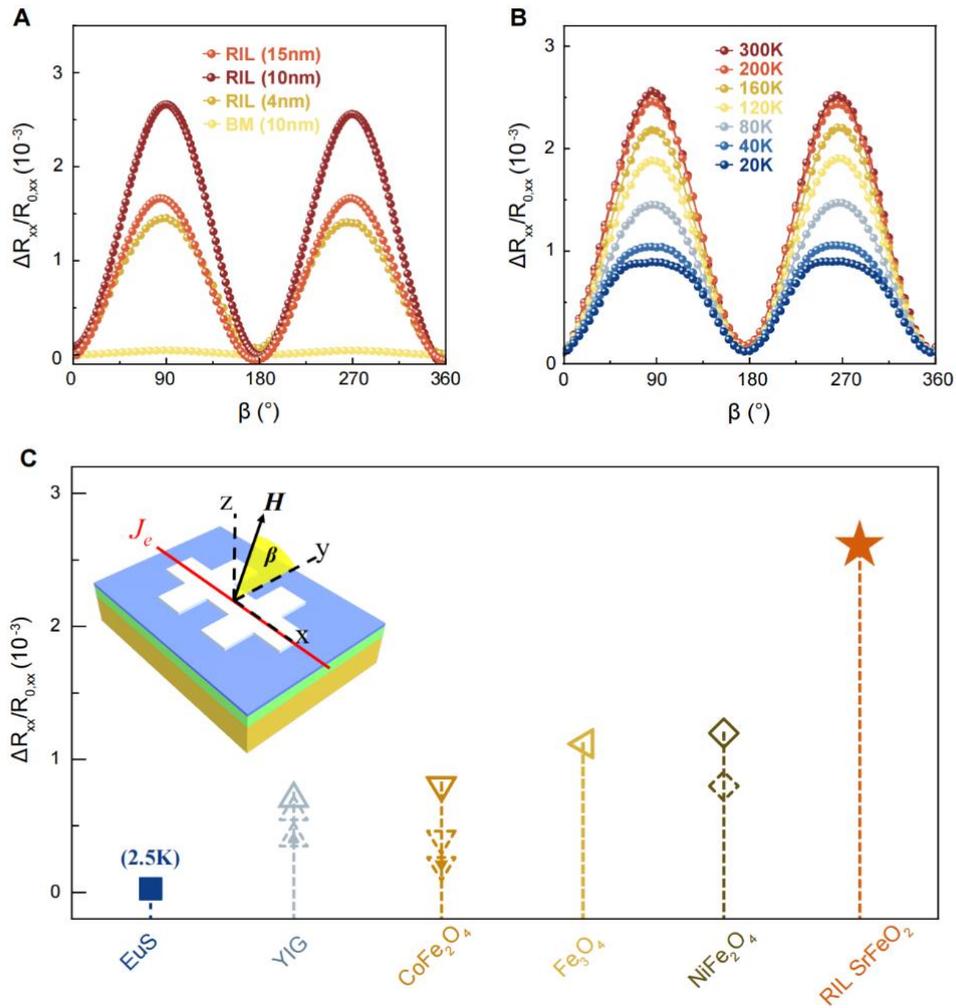

**Fig. 4.** Spin transport properties of Pt/SrFeO$_{3-\delta}$ hetero-structure systems. (**A**) Normalized longitudinal resistance of Pt/BM-SFO (10 nm), Pt/RIL-SFO (10 nm), and Pt/RIL-SFO (4 nm) systems measured at 300 K under an external magnetic field of 6 T. The numbers in parentheses indicate the thickness of the respective SFO film. (**B**) Temperature-dependent spin Hall magnetoresistance effect observed in the Pt/RIL-SFO system. (**C**) Comparative analysis of the spin Hall magnetoresistance ratio at room temperature of selected ferromagnetic (or ferrimagnetic) insulators previously reported. The inset is the schematic representation and coordinate system of the spin Hall magnetoresistance measurement setup. Note that a desirable spin Hall magnetoresistance ratio in RIL-SFO is plotted as asterisk symbols, compared to conventional ferromagnetic (solid) /ferrimagnetic (hollow) insulators (*31, 32, 44-51, 53*).